\documentclass[prl,aps,twocolumn,showpacs]{revtex4}

\usepackage[dvips]{graphicx}

\begin{document}



\title{Flow Buildup in a Mixed Phase of Quark-Gluon Plasma Plus Hadrons}

\author{Judith M. Peters and Kevin L. Haglin}

\affiliation{Department of Physics, Astronomy \& Engineering
Science, St. Cloud State University, 720 Fourth Avenue South,
St. Cloud, MN 56301 \ USA}

\date{\today}

\begin{abstract}
Transport processes are considered for their abilities to generate flow
in a mixed-phase system of hadrons plus quark-gluon plasma exhibiting 
temperature and density gradients.  Diffusive mechanisms are 
found to be the leading sources.  Even after overcoming viscous effects, 
conductive and diffusive forces generate flow velocities for pions  
in a model using kinetic theory to within fifteen percent of light speed.  
Since such flow is not observed, this could possibly rule out the mixed 
phase in the evolution of high energy heavy ion collisions.
\end{abstract}


\maketitle


The ultrarelativistic nuclear many-body problem is challenging to study
in the laboratory because system sizes are limited to roughly 
stable nuclei and the excited states are populated for about as long as 
it takes light to traverse the system.
And yet, there is a wealth of physics information to be gleaned, if
only there were a way to extract it from the collision zones created
in heavy ion collisions.  Experimental programs at the Super Proton
Synchrotron (SPS) and at the Relativistic Heavy Ion Collider (RHIC) are
rapidly advancing owing to continually developing tools for this 
extraction\cite{qm02}.
Various signals of hadronic and subhadronic degrees of freedom have been 
proposed and are being widely used.  For example, strangeness 
is expected to be enhanced and hidden charm to be suppressed at
high energy density; spacetime correlations for produced particles provide 
devices for spatial and temporal measuring; photons and dileptons
are the best temperature and spectral probes one has; and collective 
expansion as an observable could provide a dynamical measure of the equation
of state.  We study mechanisms which generate flow, and focus on the mixed 
phase.  We hope to establish the extent to which observed flow signals and 
existence of a mixed phase with gradients are inconsistent with each other.

Flow is clearly visible in momentum spectra at RHIC to a level
roughly one-half to two-thirds the speed of light depending on 
the species and the degree of centrality of the reaction\cite{phenix}.  
Such explosive 
evolution in the nuclear systems could probe salient features of underlying 
quark and gluon degrees of freedom.  The inevitable return to hadronic 
matter from highly excited quark matter might proceed through a mixed phase.
If so, and if the system has even modest temperature and density 
gradients in the mixed phase, we show that conductive and diffusive 
processes give rise to an effective mechanism for generating significant flow.
So significant that they turn out to be the leading sources; much
stronger than hydrodynamic flow from pressure gradients. 
Even after overcoming viscous effects, flow velocities 
from conduction and diffusion in a simple model reach within fifteen 
percent of light speed. 
These forces would superimpose and add to hydrodynamic forces, which
are responsible for significant flow by themselves\cite{es98,tls01}.
One could point to this as evidence ruling out the mixed phase with
a temperature gradient in the evolution of such nuclear many-body
systems.

We begin with the system in a state very near equilibrium at $T_{c}$ 
$\simeq$ 170 MeV.  Local kinetic equilibrium is assumed 
but globally we allow a temperature gradient 
$dT/dr$ $\simeq$ $-$1 MeV/fm.  This is the simplest assumption,
although other profiles could be envisaged.  We note that
a temperature gradient is not inconsistent with hydrodynamics since
a local maximum temperature different by as much as 70 MeV from
the average temperature has been estimated\cite{hrr02}.
By this stage in the evolution of the 
nuclear collisions, 
probably 5--10 fm/$c$ after maximum overlap, the system is roughly spherical 
in shape with a radius of $d$ = 10 fm.  For definiteness, let us indicate 
temperatures are central temperatures, and we here assume zero skin 
thickness.  Two quark flavors plus gluons and prehadrons (forming
hadrons) are assumed to be in local thermal equilibrium.
In practice, this corresponds to densities of 5.5, 0.30, and 0.08 per 
fm$^{3}$ for massless partons, pions and protons, respectively.
It has been recently pointed out that under these
conditions thermophoresis is responsible for generating
significant flow\cite{mt02}.

Description of irreversible phenomena in systems displaced somewhat
from equilibrium are dominated by thermodynamic forces
and flows.   The flows tend to smooth out the inhomogeneities
introduced by the forces.  One such example is heat conduction which
is responsible for a force per unit area on each hadron given by 
\begin{eqnarray}
F/A & = & -\,{1\over\/3}\,n\,\langle\/u\/\rangle\,
\lambda\,{d\,\langle\/E\,\rangle\over\/d\/T\/}\,{d\/T\over\/d\/r\/}\,,
\label{foacond}
\end{eqnarray}
where $n$ is the number density of partons in the quark-gluon plasma, 
$\langle\/u\/\rangle$ is the mean parton velocity, $\lambda$ their mean 
free path and $\langle\/E\,\rangle$ is their average energy.
The net thermophoretic force (from conduction) acting on the hadrons due to
parton dynamics is then given by
\begin{eqnarray}
F & = & \pi\/R^{\/2}\left(F/A\right)\,,
\end{eqnarray}
where $R$ is the hadron radius.  Using elementary methods, one
can estimate from here the result of this radially outward force
in terms of the asymptotic hadron velocity expected in a
picture relevant for high-energy heavy-ion collisions.  In \cite{mt02} Thoma 
found the final radial velocity for pions to be $v_{f}$ = 0.81 and
for protons $v_{f}$ = 0.40, although see~\cite{correctionI} for
a correction. 

We go beyond the effects of conduction and begin to study
other transport processes in the presence of the same temperature
gradient.  If there were a temperature gradient, then the 
default scenario is to support a density gradient as well.  Particles
will of course diffuse from regions of high density to regions of lower
density.  Parton diffusion, we argue, is therefore responsible
for a force per unit area 
\begin{eqnarray}
F/A & = & 
-\,{1\over\/3}\,\langle\/u\/\rangle\, \lambda\,\langle\,p\,\rangle
{d\,n\,\over\/d\/T\/}\,{d\/T\over\/d\/r\/}\,,
\label{foadiff}
\end{eqnarray}
where this time the spatial variation in density drives the
mechanism.  The average momentum appears rather than energy 
trivially, since force is proportional to $\dot{p}\,$\cite{reif}.

Finally, the last effect we include is viscosity.  Viscous
effects are notoriously difficult to include in hydrodynamics, but
here we use elementary kinetic theory argumentation to write
\begin{eqnarray}
F/A & = & {1\over\/10}\,n\,\langle\/u\/\rangle\,
\lambda\,{d\langle\,p\,\rangle\over\/d\/T\/}\,{d\/T\over\/d\/r\/}\,.
\label{foavisc}
\end{eqnarray}
Viscosity effects, like diffusion, involve average momentum
rather than energy, as was the case for conduction which
involves energy transport.  In the massless limit the distinction disappears.
Notice also that viscosity in this approach is 3/10 of 
the conductive force and that the direction
of the viscous force opposes conduction and diffusion\cite{prefactor}.
Instead of using the simplified approach taken here, one could compute
collision brackets in the first Chapman--Enskog approximation
for linearizing the full parton kinetic theory\cite{dvv}.  However,
the elegant results could get lost in the details.
We also note that viscosity of a quark-gluon plasma
has been estimated previously using transport theory\cite{bmpr90}
as well as kinetic theory\cite{mt91}.

Now we explore two different pictures: 1) a system of massless partons
plus hadrons and, 2) a system of partons with dynamically-generated 
masses plus hadrons, with $m_{q}$ = $g\/T/\sqrt{6}$ for quarks and 
antiquarks and $m_{g}$ = $g\/T/2\left(1+N_{f}/6\right)$ for gluons, where
$N_{f}$ is the number of flavors\cite{ml,lr1}.  The idea is to compare
the preliminary and rough estimates of conduction by Thoma to the 
more quantitative formalism discussed here, to compare also with 
diffusion, and finally, to explore finite-mass effects arising from the 
medium.

It is particularly convenient to first consider massless partons since the
expressions can then be reported very compactly.  To set the notation 
and indicate explicitly the partonic states
considered, let $g_{q}$ = $g_{\:\bar{q}}$ = $N_{s}\times\/N_{c}\times\/N_{f}$
= 2$\times$3$\times$2 = 12
and $g_{g}$ = $N_{s}\/\times\/N_{c}$ = 2$\times$8 = 16 for the 
degeneracies.  Noting that
$\lambda_{q}$ = $\lambda_{\bar{q}}$ = 4/9$\lambda_{g}$, we find 
\begin{eqnarray}
\left.{F/A\/}\right)_{\rm\/cond} & = & -{T^{3}\over\pi^{2}}\,\zeta(4)\,
\lambda_{q}\left\lbrack{7\over\/4}g_{q}+{4\over\/9}g_{g}\right\rbrack\,
{dT\/\over\/dr\/}\,,\nonumber\\
\left.{F/A\/}\right)_{\rm\/diff} & = & -{T^{3}\over\/\pi^{2}}\,\zeta(4)\,
\lambda_{q}\left\lbrack{21\over\/4}g_{q}+{4\over\/3}g_{g}\right\rbrack\,
{dT\/\over\/dr\/}\,,\nonumber\\
\left.{F/A\/}\right)_{\rm\/visc} & = & {T^{3}\over\/\pi^{2}}\,\zeta(4)\,
\lambda_{q}\left\lbrack{21\over\/40}g_{q}+{2\over\/15}g_{g}\right\rbrack\,
{dT\/\over\/dr\/}\,.
\end{eqnarray}
Since $\zeta(4)$ = $\pi^{4}/90$~\cite{as}, and since
$\lambda_{q}$ = [3$C_{F}\alpha_{s}\,T\/]^{-1}$ (with the Casmir
invariant $C_{F}$ = 4/3) has been computed in studies of
parton relaxation rates using a hard-thermal-loop (HTL) 
approximation\cite{mt93}, and finally plugging in the degeneracies, we 
arrive at
\begin{eqnarray}
\left.{F/A\/}\right)_{\rm\/cond} & = & -{\pi^{2}\,T^{2}\over\/4\alpha_{s}}
\left({253\over\/810}\right){dT\/\over\/dr\/}\,,\nonumber\\
\left.{F/A\/}\right)_{\rm\/diff} & = & -{\pi^{2}\,T^{2}\over\/4\alpha_{s}}
\left({253\over\/270}\right){dT\/\over\/dr\/}\,,\nonumber\\
\left.{F/A\/}\right)_{\rm\/visc} & = & {\pi^{2}\,T^{2}\over\/4\alpha_{s}}
\left({253\over\/2700}\right){dT\/\over\/dr\/}\,.
\end{eqnarray}
For simplicity, we assume spherical hadrons (prehadrons, or 
extended hadron bubbles) of radius $R \simeq$ 1 fm and
compute $F = \pi\/R^{2}\left(F/A\right)$ for the three 
transport mechanisms.
\begin{figure}[t]
\begin{center}
\includegraphics[width=6.0cm,height=7.5cm]{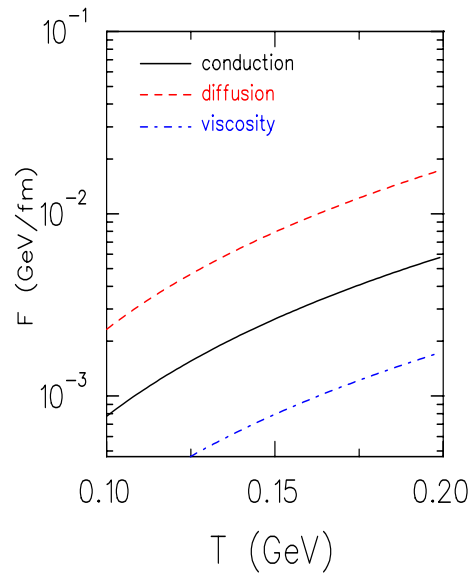}
\caption{Magnitudes of forces due to a temperature gradient. 
Heat conduction effects (solid curve), diffusion (dashed curve) and 
viscosity effects (dot-dashed curve) are shown separately.}
\label{forcesI}
\end{center}
\end{figure}
The thermodynamic forces turn out to be proportional to the square of the 
temperature times the temperature gradient.
However, we allow the strong coupling constant to run
with temperature and approach an asymptotically free value in 
the $T\to\infty$ limit according to\cite{lr2}
\begin{eqnarray}
\alpha_{s}(\/T\/) & = &
{\alpha_{s}(T_{c})\over\/1+\/C\,\ln\left(T/T_{c}\right)}\,,
\label{alphastrong}
\end{eqnarray}
where $C \simeq$ 0.76, $\alpha_{s}(T_{c})$ = 0.48,
and $T_{c}$ = 170 MeV.
The final dependence of the forces on temperature is softer than
$T^{2}$ as shown in Fig.~\ref{forcesI}.
The results suggest that diffusion seems to be responsible for the 
dominant force.

These radial forces act on prehadrons of mass $M$ and consequently generate
radial flow.  The force is constant so quite simply
the final hadron kinetic energy is equal to the net force times
the distance $d$, {\it\/i.e.\/}
$K$ = $(F_{\rm\/cond}+F_{\rm\/diff}+F_{\rm\/visc})\,d\/$ (since
$F_{\rm\/visc} < 0$).
Using relativistic kinematic formul\ae, we find\cite{correctionI}
\begin{eqnarray}
\beta_{f} & = &  
\sqrt{1-\left({M\over\/K+M\/}\right)^{2}}\,.
\label{vfinal}
\end{eqnarray}
\begin{figure}[t]
\begin{center}
\includegraphics[width=6.0cm,height=7.5cm]{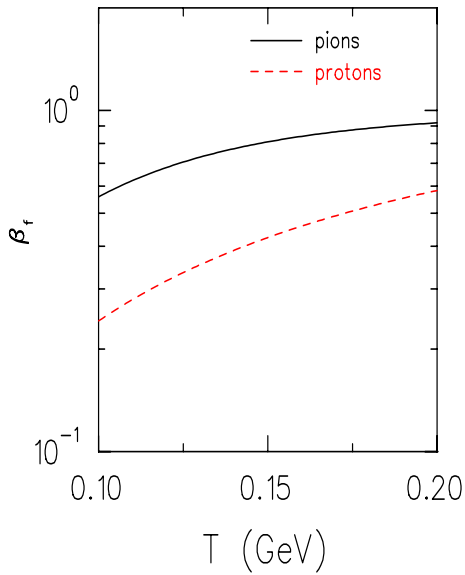}
\caption{Final flow velocities for pions (solid curve) and protons 
(dashed curve) generated by transport processes.}
\label{vfinalI}
\end{center}
\end{figure}
The final flow velocities for pions and protons using
$d$ = 10 fm are plotted in 
Fig.~\ref{vfinalI} as a function of temperature.  At $T = T_{c}$ = 170 MeV
the values are 0.86 and 0.49, respectively.

\begin{figure}[t]
\begin{center}
\includegraphics[width=6.0cm,height=7.5cm]{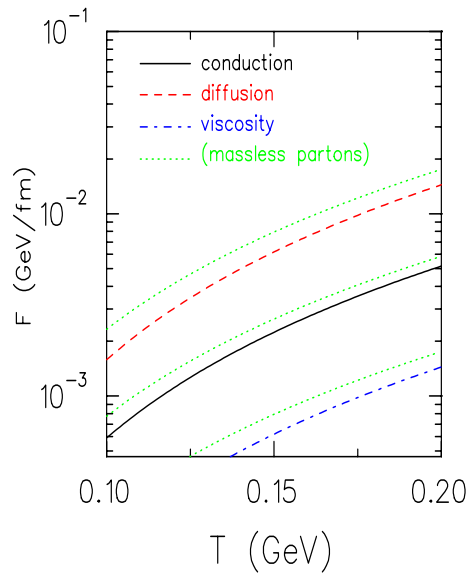}
\caption{Transport force magnitudes from massive quarks
and gluons.  Conduction is shown as the solid curve, diffusion 
is the dashed curve and viscosity is shown in the dot-dashed
curve.  For each curve, the massless parton limit is plotted
as the dotted curve immediately above.}
\label{forcesII}
\end{center}
\end{figure}

Next we explore the first-order effects of the medium to the extent that
thermal masses could change these results.  Relevant quantities
appear in the form
\begin{eqnarray}
\langle\,{\cal\/O\,}\rangle & = & 
\left.{\int\,d\/^{3}p\,\left[{\cal\,O\,}\right]f(p_{0})}\right/
{\int\,d^{3}p\,f(p_{0})}\,,
\end{eqnarray}
where $p_{0} = \sqrt{\vec{p\/}^{\,2}+m^{\,2\/}}$ and
where equilibrium distributions
$f$ = 1/\,$\left[\/\exp\,(\beta\/p_{0})\pm\/1\right]$ 
are taken as appropriate for Bosons (Fermions) using the
$-$ ($+$) sign.   We calculate
\begin{eqnarray}
n & = & {g\,T^{\,3}\over\/2\/\pi^{2}\/}
\sum\limits_{\ell=1}^{\infty} {\left(\pm\right)^{\ell\,+1}\over\ell}
\,x^{\,2}\,K_{2}\left(x\,\ell\,\right)
\nonumber\\
\langle\,p\,\rangle & = & {g\,T^{4}\over\pi^{2}\,n\/}
\sum\limits_{\ell=1}^{\infty} {\left(\pm\right)^{\ell\,+1}\over\ell^{4}}
\left\lbrack{x^{\/2}\ell^{\,2}}
+{3\,x\,\ell}
+3\right\rbrack\nonumber\\
& \ & \times\,e^{-x\,\ell\,}\,
\nonumber\\
\langle\,E\,\rangle & = & {g\,T^{4}\over\,2\,\pi^{2}\,n\/}
\sum\limits_{\ell=1}^{\infty} {\left(\pm\right)^{\ell\,+1}\over\ell^{2}}
\left\lbrack
{x^{\,3}\,\ell}K_{1}\left(\,x\,\ell\,\right)\right.
\nonumber\\
& \ & \left.
+{3\,x^{\,2}\/}
K_{2}\left(x\,\ell\,\right)
\right\rbrack
\nonumber\\
\langle\,u\,\rangle & = & {g\/T^{3}\over\/\pi^{2}\,n\/}
\sum\limits_{\ell=1}^{\infty} {\left(\pm\right)^{\ell\,+1}\over\ell^{3}}
\left( {x\,\ell}\,+\,1\right)\,e^{-x\,\ell\,}\,,
\label{avgvalues}
\end{eqnarray}
where $x$ $\equiv$ $m\,\beta$ = $m/T$,
$K_{i}$ is the modified Bessel function of order $i$, 
$g$ is the appropriate degeneracy, $n$ is the density 
of quarks or gluons, 
$\langle\,u\,\rangle$ is the average speed and the choice 
of sign in the summands corresponds to $+$ for gluons (Bosons) and 
$-$ for quarks (Fermions).
The massless limits for the above expressions are 
naturally recovered when $x\to\/0$ since
\begin{eqnarray}
\lim\limits_{z\to\/0}\,\left(z\/K_{1}(z)\right) 
& = & 1
\nonumber\\
\lim\limits_{z\to\/0}\,\left(z^{\,2}\/
K_{2}(z)\right) 
& = & 2\,.
\end{eqnarray}
In that case $\langle\,p\,\rangle$ = $\langle\,E\,\rangle$ and
$\langle\,u\,\rangle$ = 1.  Truncation of the sums using only $\ell$ = 1
terms corresponds to the Boltzmann limit.  Temperature derivatives
of the quantities in Eq.~(\ref{avgvalues}) can be readily computed
as needed.

The first observation regarding the thermal masses is that the
effects on forces shown in Fig.~\ref{forcesII} are quite small (at the 
ten percent level).  This insensitivity to medium effects 
on the forces and the expected flow is consistent with previous 
findings\cite{cr99}.  The effects are slightly more apparent at lower
temperatures mainly due to the density.

\begin{figure}[t]
\begin{center}
\includegraphics[width=6.0cm,height=7.5cm]{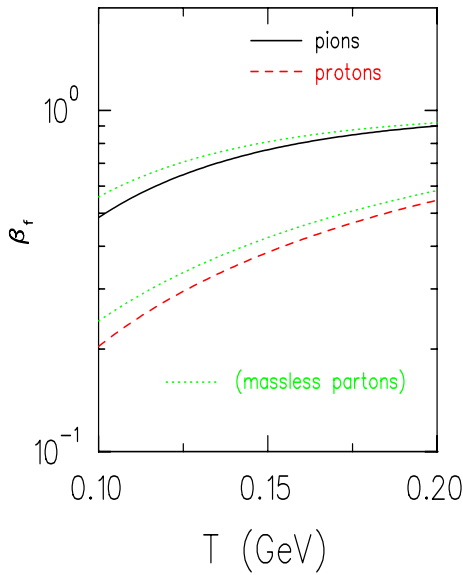}
\caption{Final flow velocities due to thermal-parton transport forces
for pion (solid curve) and protons (dashed curve).  Results for 
massless partons are also shown for each case (dotted curves).}
\label{vfinalII}
\end{center}
\end{figure}

For a symmetrical presentation of the results, we include 
Fig.~\ref{vfinalII}
which reports the final flow velocities of hadrons (pions and
protons) due to transport forces from massive partons.  Results
using massless partons obtained previously are shown as well
in the figure.

Before concluding, a remark is in order regarding the choice of gradient 
and system size.  Since the final hadron kinetic energy is equal to the 
product of the temperature gradient times the system size $d$, the same 
results would emerge if one instead took $d$ = 1 fm, for then 
$dT/dr$ would be $-$10 MeV/fm.   Consequently, the expected
flow would be unchanged.

We summarize our study as follows.
Transport processes have been studied in a mixed phase of quarks and
gluons plus hadrons at or near $T_{c}$.  A modest temperature gradient
is responsible for a net radial force on the hadrons having magnitude
11 MeV/fm.  The leading mechanism or source of
flow is parton diffusion which pushes hadrons radially outward
roughly 9 MeV/fm.  Viscosity introduces a drag force of $\sim$ 1
MeV/fm.  In a simple picture, these forces are responsible for
flow velocities 0.49$c$ for protons and 0.86$c$
for pions.  The first-order effect of the medium was included,
in which case the forces dropped by 10--20 percent.  Flow velocities
were consequently scaled back slightly as well, although relativistic 
velocity dependences on forces are clearly nonlinear.

One might suggest that these result argue for the nonexistence of
the mixed phase, since flow velocities greater than 0.6$c$ are not
seen in the data at RHIC.  Instead, the expansion and evolution of
the fireball in heavy ion collisions could proceed very quickly through 
the phase boundary and smoothly as in say, a second order
(or higher) phase transition from quark-gluon plasma to hadronic
matter.

{\bf Acknowledgment.}
We thank Markus H. Thoma for useful discussions.
This work was supported in part by the National Science Foundation
under grant number PHY-0098760.



\begin{thebibliography}{99}
\bibitem{prefactor}The coefficients out in front of 
Eqs.~(\ref{foacond}) and (\ref{foavisc}) are not rigorously identified 
here.  The best one can really do without a proper calculation of the
transport coefficients is to indicate that the prefactors are of order 1.
\bibitem{correctionI}Comparing the development here to Eq.~(3) in 
Ref.~\cite{mt02}, we notice that 
an incorrect assumption was implemented there converting the derivative from 
a time to a spatial derivative. The erroneous assumption is a constant 
velocity.  Therefore, Eqs.~(4) and (5) of Ref.~\cite{mt02} are slightly 
incorrect and our Eq.~(\ref{vfinal}) is the correct one.
\end{thebibliography}
\end{document}